\documentclass[twocolumn,prl,superscriptaddress]{revtex4}
\newcommand{\pagenumbaa}{1}
\usepackage{ulem}

\usepackage{amsmath}
\usepackage{graphicx}
\usepackage{setspace}
\usepackage{units}
\usepackage{color}
\usepackage{hyperref}

\newcommand{\be}{\begin{equation}}
\newcommand{\ee}{\end{equation}}
\begin{document}

\title{Spectroscopy of Quantum-Dot Orbitals with In-Plane Magnetic Fields}

\author{Leon C. Camenzind}
\affiliation{Department of Physics, University of Basel, Basel 4056, Switzerland}

\author{Liuqi Yu}
\affiliation{Department of Physics, University of Basel, Basel 4056, Switzerland}

\author{Peter Stano}
\affiliation{Center for Emergent Matter Science, RIKEN, Saitama 351-0198, Japan}
\affiliation{Department of Applied Physics, School of Engineering, University of Tokyo, 7-3-1 Hongo, Bunkyo-ku, Tokyo 113-8656, Japan}
\affiliation{Institute of Physics, Slovak Academy of Sciences, 845 11 Bratislava, Slovakia}

\author{Jeramy Zimmerman}
\affiliation{Materials Department, University of California, Santa Barbara 93106, USA}

\author{Arthur C. Gossard}
\affiliation{Materials Department, University of California, Santa Barbara 93106, USA}

\author{Daniel Loss}
\affiliation{Department of Physics, University of Basel, Basel 4056, Switzerland}
\affiliation{Center for Emergent Matter Science, RIKEN, Saitama 351-0198, Japan}

\author{Dominik M. Zumb\"uhl}
\affiliation{Department of Physics, University of Basel, Basel 4056, Switzerland}

\setcounter{page}{\pagenumbaa}
\thispagestyle{plain}

\begin{abstract}
We show that in-plane-magnetic-field assisted spectroscopy allows extraction of the in-plane orientation and full 3D shape of the quantum mechanical orbitals of a single electron GaAs lateral quantum dot with sub-nm precision. The method is based on measuring orbital energies in a magnetic field with various strengths and orientations in the plane of the 2D electron gas. As a result, we deduce the microscopic quantum dot confinement potential landscape, and quantify the degree by which it differs from a harmonic oscillator potential. The spectroscopy is used to validate shape manipulation with gate voltages, agreeing with expectations from the gate layout. Our measurements demonstrate a versatile tool for quantum dots with one dominant axis of strong confinement.
\end{abstract}
\maketitle

A spin in a magnetic field is one of the simplest canonical quantum two-level systems encoding a qubit \cite{Loss1998}. To realize spin based quantum computing, the capability of addressing individual spin qubits is essential, as demonstrated in various semiconductor quantum dot devices \cite{Kloeffel2013}. Although significant progress has been made on the control of spin states, a comprehensive tool to tune and characterize the quantum dot---the spin qubit host---is still missing. This is one of the key factors for qubit performance, but the challenge lies in the lack of means to adjust the confinement potential, particularly for dot systems formed in nanowires or by intrinsic defects. Lateral quantum dots, on the other hand, show excellent flexibility. Defined in a 2D electron gas (2DEG) by nanometer-scale surface gates, they allow, in principle, arbitrary and tunable dot shapes and orientations \cite{Amasha2008}.

This tunability provides an additional knob important for stabilizing and manipulating the spin states \cite{Malkoc2016}. The dot shape, impressed in its orbital energy spectrum, is directly associated with a variety of spin-electric related processes. These rely on mixing of spin and orbital degrees of freedom since the orbital shape determines the dipole moments connected with spin-flip transitions. For instance, such mixing presents the predominant channel for spin relaxation, in GaAs through both the spin-orbit \cite{Amasha2008,Scarlino2014, Golovach2004}, and hyperfine interactions \cite{Camenzind2017,Erlingsson2002}. Both spin relaxation \cite{Amasha2008} and spin manipulation by electric-dipole spin resonance (EDSR) \cite{Nowack2007, Pioro-Ladriere2008} show a strong dependence on the dot shape and the orientation in the 2DEG plane and with respect to the magnetic field. The dependence can be exploited to control both the spin relaxation time and EDSR frequency \cite{Malkoc2016}.

The bottleneck in taking full advantage of this flexibility is that so far there is no direct method to determine the quantum-dot shape. Since the dot is imprinted into the 2DEG beneath the surface of the device, details of the dot shape are inaccessible for surface imaging tools, such as atomic force or scanning tunneling microscopy. In addition, the electric fields from the surface gates will in return interfere with the probe aggravating such measurements \cite{Topinka2001,Pioda2004}. Further, these methods suffer from invasive back-action of the probe to the sample disturbing the quantum dot. In principle, advanced simulation software allows to precisely calculate the potential profile. However, all such simulations rely solely on the design of the nanogates on the surface. They do not capture all realistic conditions in the device, such as fields from charge impurities or the strain distribution. These sample-dependent and largely unknown factors can strongly influence the dot shape and position, making it often different from the geometry suggested by the gate layout alone. The above methods can therefore serve as a rough guide, but are unable to provide sub-dot resolution of the confinement potential.

In this Letter, we present a non-invasive technique which is able to reconstruct the shape and orientation of quantum mechanical orbitals of a quantum dot. It is based on measuring the orbital energy spectrum dependence on in-plane magnetic field of varying magnitude and direction. We show that this technique allows to resolve the full 3D confinement of a single-electron quantum dot with sub-nm precision. The theoretical principles of the method are explained in Ref.~\citenum{Stano2017}. Here, we demonstrate it experimentally.

\begin{figure}
\center
\includegraphics[width=0.95\columnwidth]{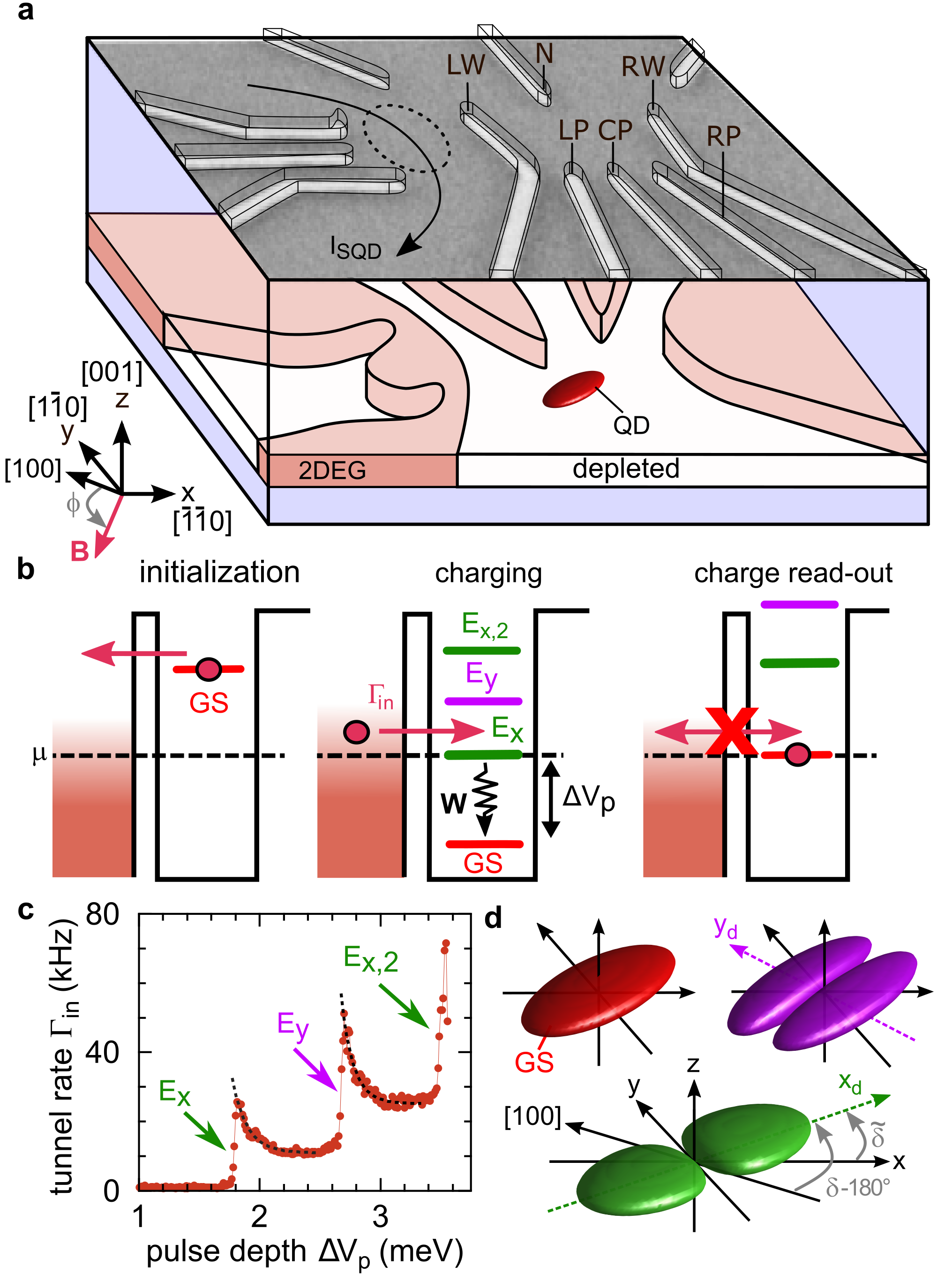}\vspace{-2mm}
\caption{(a) Sketch with electron micrograph of the gate layout of a cofabricated device shown on top. The GaAs/ Al$_{0.3}$Ga$_{0.7}$As heterostructure contains a 2DEG with density $2.6\times10^{11}\,\text{cm}^{-2}$ and mobility $4\times10^5\,\text{cm}^2/\text{Vs}$ located 110\,nm below the surface. The in-plane field angle $\phi$ and wave function orientation $\delta$ are defined with respect to $[100]$, while $\tilde{\delta}=\delta-225^\circ$ is the angle between $\hat{x}=[\bar{1}\bar{1}0]$ and $\hat{x}_{d}$, the dot confinement x-axis.
(b) Three-step pulse sequence described in the text.
(c) Measurement of tunneling-in rate $\Gamma_{in}$ as a function of $\Delta V_{P}$ exhibiting three excited orbital states at energies $E_x$, $E_y$ and $E_{x,2}$.
(d) Ground-state wave function (left) and the p-type orbitals for an elongated dot with exaggerated anisotropy, details in \cite{Supp}S1. }
\vspace{-4mm}
\label{fig:Fig1}
\end{figure}

The surface gate layout of the measured device, shown in Fig.~\ref{fig:Fig1}(a), is based on Ref.~\onlinecite{Amasha2008}. The device is biased into the single-electron quantum-dot regime, as indicated by the red ellipsoid. The dot is tuned to couple to the left reservoir only, with a tunnel rate between $1$\, and $100$\,Hz. An additional quantum dot, located directly adjacent to the main dot, is serving as a charge sensor \cite{Field1993,Barthel2010}, giving a change of sensor conductance of up to 100\% per electron in the main dot. The sample can be oriented with essentially arbitrary angle with respect to an in-plane magnetic field up to $14$\,T using a piezoelectric rotator. Using standard van der Pauw measurements, the magnetic field is shown to deviate less than $1.3^{\circ}$ out of the 2DEG plane, thus rendering the out-of-plane component negligible \cite{Camenzind2017}. Measurements are done in a dilution refrigerator with an electron temperature of $60$ mK \cite{SchellerAPL2014, Maradan2014,Biesinger2015}.

The orbital energies are measured by pulsed gate spectroscopy using a three-step pulse sequence. Namely, an additional voltage $\Delta V_{P}$ is applied to the center plunger gate $CP$, on top of the static gate voltage $V_{P}$, see Fig.~\ref{fig:Fig1}(a) \cite{Elzerman2004,Johnson2005,Amasha2008}. As illustrated in Fig~\ref{fig:Fig1}(b), the sequence consists of initialization, charging and read-out steps, see also \cite{Supp}S5. Compared to the ground state, the elastic tunnel rate into the empty dot increases sharply when an excited orbital state becomes resonant with the chemical potential $\mu$ of the reservoir. By measuring the dot-reservoir tunnel coupling for varying $\Delta V_{P}$, individual excited orbital states can be distinguished. An example is shown in Fig.~\ref{fig:Fig1}(c) exhibiting three excited orbital states. The ground state, which calibrates $\Delta V_{P} = 0$, couples much weaker to the reservoir ($\Gamma_{GS} \sim 10$\,Hz) compared to the excited orbital states, attributed to the increased spatial extent of higher orbitals \cite{Fujisawa2001,Hanson2003a}. The exponential decay in the tunnel rate of the excited states with increasing $\Delta V_{P}$ [dashed curves in Fig.~\ref{fig:Fig1}(c)] is associated with an increasing tunnel barrier experienced by an electron tunneling into the empty quantum dot as the gate voltage is changed \cite{MacLean2007, Amasha2008b, Stano2010a}.

A 2D harmonic oscillator model is used to describe the in-plane confinement of the quantum dot while a much stronger confinement $v(z)$ is taken along the growth direction $\hat{z}$ due to the heterostructure interface:
\begin{equation}
\label{eq:H}
H = \frac{\textbf{p}^2}{2m} + \frac{\hbar^2}{2m}\left( \frac{x_d^2}{l_x^4}+\frac{y_d^2}{l_y^4} \right) + v(z).
\end{equation}
Here, $\textbf{p}$ is the momentum operator, $\hbar$ the reduced Planck constant, $m$ the effective mass, and $l_{x,y}$ are the confinement lengths along the main axes $\hat{x}_d$ and $\hat{y}_d$ of the in-plane confinement. These axes are in general rotated from the crystal axes [100] and [010] by an angle $\delta$, see Fig.~\ref{fig:Fig1}(a,d). For simplicity, we introduce $\tilde{\delta}=\delta-225^\circ$ as the angle between wave function axis $\hat{x}_d$ and device axis $\hat{x}=[\bar{1}\bar{1}0]$. In the model described by Eq.~\eqref{eq:H}, the excitation energies are $E_{x,y} = \hbar^2/m l_{x,y}^2$ and the ground-state wave function can be represented by a disk-like ellipsoid. The two lowest excited states correspond to p-like orbitals aligned along two perpendicular axes $\hat{x}_d$, $\hat{y}_d$, as shown in  Fig.~\ref{fig:Fig1}(d).

\begin{figure}\hspace*{-3mm}
\includegraphics[width=1.0\columnwidth]{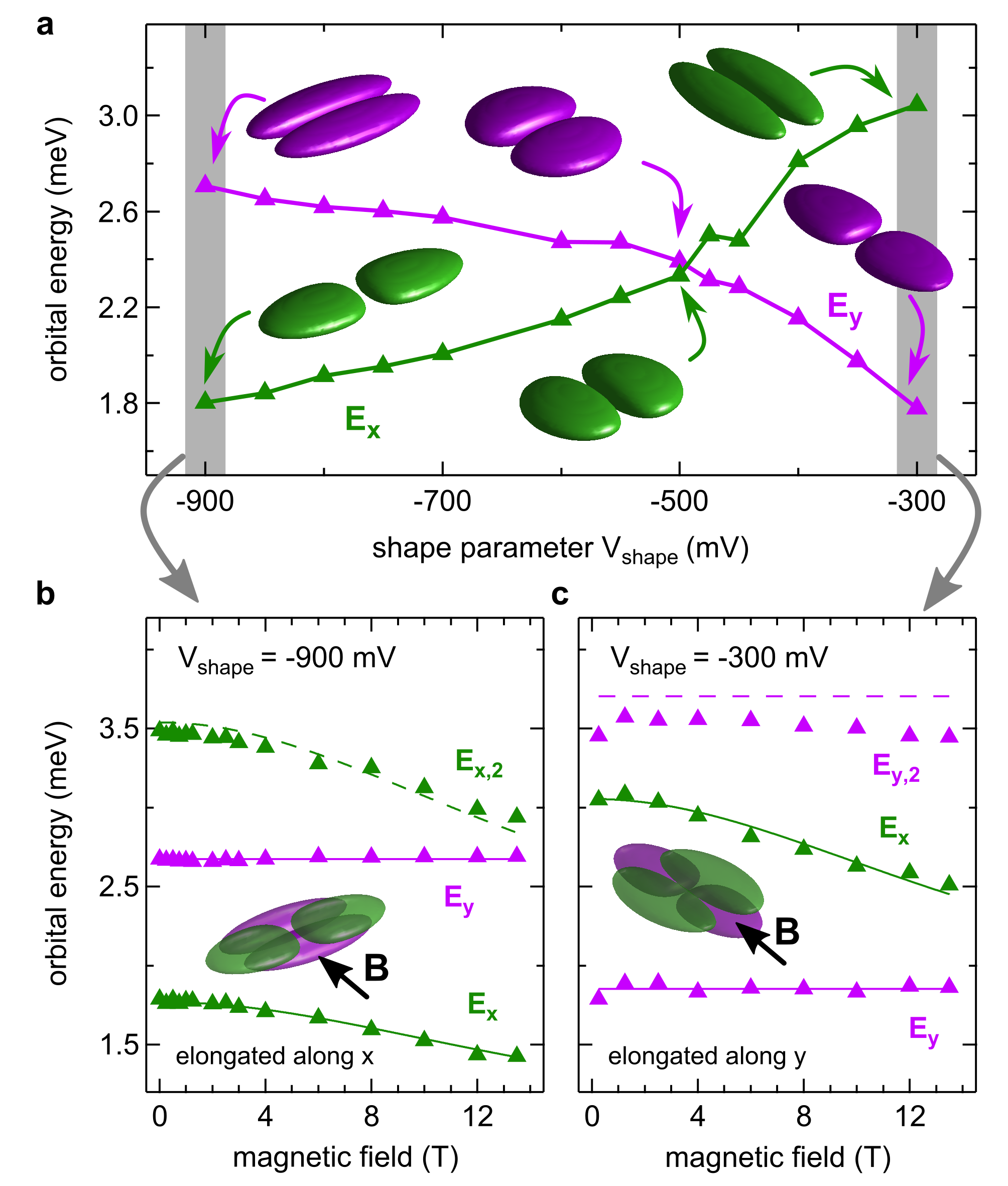}\vspace{-2mm}
\caption{(a) Orbital excitation energies $E_x$ (green) and $E_y$ (purple) as a function of $V_{\rm shape}$. The schematics give a qualitative picture of the excited orbital wave functions along the $\hat{x}$-direction (green) and $\hat{y}$-direction (purple) for the three shapes indicated by the arrows. Less exaggerated wave functions are shown in \cite{Supp}S1. Orbital excitation energies for a magnetic field applied along the $y$-direction for two extreme dot shapes (b) elongated along $\hat{x}$ and (c) elongated along $\hat{y}$, for $V_{\rm shape}$ as labeled. The data are fitted to Eq.~\eqref{eq:dE}, giving $\lambda_z = 6.3\pm0.3$\,nm and $E_z = 28.6\pm{3}$\,meV \cite{GeneralizedFit}.\vspace{-5mm}}
\label{fig:Fig2}
\end{figure}

Within this model, the parameters $E_x$, $E_y$, and $\delta$ characterize the dot shape, and vice versa, control of these parameters indicates dot-shape tunability. This is what we demonstrate next. Applying appropriate voltages on the surface gates, the dot can be elongated either in the $\hat{x}$- or alternatively in the $\hat{y}$-direction \cite{Amasha2008}. For instance, the dot can be squeezed in the $\hat{y}$-direction by applying more negative voltages on the plunger gates LP, CP and RP, see Fig.~\ref{fig:Fig1}(a). To keep the ground-state energy constant, these changes are compensated by applying less negative voltages on the other gates LW and RW, which leads to an expansion of the wave function in the $\hat{x}$-direction. We introduce a shape parameter $V_{\rm shape}$ to denote the full set of gate voltages corresponding to a particular dot shape (see Fig.~\ref{fig:Fig2}), with the numerical value of $V_{\rm shape}$ taken to be the voltage on gate CP.

The two lowest orbital excitation energies are shown in Fig.~\ref{fig:Fig2}(a) as a function of the dot shape $V_{\rm shape}$. Upon making $V_{\rm shape}$ more negative, thus squeezing the dot in the $\hat{y}$-direction, one of the two energies increases, indeed, which thus is identified as the $\hat{y}$ state. The other energy decreases, and thus has to be the $\hat{x}$ state, as labeled in Fig.~\ref{fig:Fig2}. Interestingly, at $V_{\rm shape} \sim -500$\,mV, we find $E_{x}\approx E_{y}$, indicating a circular, isotropic wave function in the 2D plane. Such shape manipulation by gate voltages is limited on one hand by the minimum voltage needed to deplete the 2DEG underneath the surface gates, and on the other hand by the gate leakage threshold at more negative gate voltages. We emphasize that throughout the shape manipulation, the tunneling rate to the reservoir is held approximately constant. For each dot shape, the relevant lever arm is measured, providing the gate voltage to energy conversion in order to obtain the excited state energies from pulsed gate spectroscopy, see \cite{Supp}S4 for details.

From such data, however, there is no estimate of the tilt angle $\tilde{\delta}$ or how it depends on $V_{\rm shape}$---other than that it is probably not too big. It is natural to expect that, as the dot is being squeezed, the wave function is also shifted and possibly somewhat rotated in space, depending on the detailed potential and disorder landscape present. In addition, we note that the subband excitations $E_{z} \gg E_{x,y}$ are energetically out of reach of this pulsed-gate spectroscopy method, so that little can be said about the size of the dot orbitals along the growth axis. We are now going to show how this missing information can be revealed; this is the main advance that our work makes.

To this end, we exploit the effects of a strong in-plane magnetic field $\mathbf{B}$ applied along an in-plane direction $\hat{b}$, which makes an angle $\phi$ with the [100] crystallographic axis, see coordinate system in Fig.~\ref{fig:Fig1}(a). In Ref.~\cite{Stano2017}, we show that the leading order effect can be expressed as a correction to Eq.~\eqref{eq:H} of the following form \cite{Stern1968}
\begin{equation}
\delta H = -\frac{\Phi^2}{2m} \left[\textbf{p}\cdot\left( \hat{b} \times \hat{z} \right) \right]^2.
\label{eq:dH}
\end{equation}
This interaction is the basis for our spectroscopy. Its strength scales with the magnetic flux $\Phi$ penetrating the 2DEG due to its finite width. Explicitly,
\begin{equation}
	\Phi = \frac{e}{\hbar}B\lambda_z^2,
\label{eq:flux}
\end{equation}
where $e>0$ is the elementary charge and $\lambda_z$ is the effective width of the wave function along the growth direction. We analyze the connection between a nominal width and the effective width of a 2DEG for several confinement profiles, namely triangular, harmonic, and a square potential well \cite{Stano2017}. Also, we note that flux threading was previously studied in open dots \cite{Falko2002,ZumbuhlPRB2004,ZumbuhlPRB2005}.

For typical 2DEGs and magnetic fields, the flux is small: $\Phi\ll 1$ \cite{GeneralizedFit}.
Treating Eq.~\eqref{eq:dH} as a perturbation to Eq.~\eqref{eq:H}, the energies change by
\begin{equation}
	\delta E_{x,y} = -\frac{\Phi^2}{2} \frac{\hbar^2}{m l_{x,y}^2} \sin^2(\delta_{x,y} - \phi).
\label{eq:dE}
\end{equation}
Here, we denoted $\delta_{x,y}$ as the corresponding excited orbital directions (with respect to [100]). They follow from Eq.~\eqref{eq:H} as $\delta_x = \delta$ and $\delta_y = \delta + \pi/2$.

First, we apply a strong magnetic field along the $y$-direction for the two most elongated shapes available, see Fig.~\ref{fig:Fig2}(b) and (c). For sufficiently weak confinement along one direction, a second excited state $E_{\rm x,2}$ or $E_{\rm y,2}$ also becomes accessible. While $E_{x,2}\sim 2 E_x$ for the dot in Fig.~\ref{fig:Fig2}(b), $E_{y,2}$ is slightly lower in energy than the second harmonic of $E_y$, as seen in Fig.~\ref{fig:Fig2}(c). For this configuration, the voltage on the nose N and all plunger gates are only barely sufficient to deplete the 2DEG, which could lead to a softening of the confinement potential along $\hat{y}$. Looking at the field dependence, we make the striking observation that $E_{y}$ remains constant for both shapes while $E_x$ clearly changes with magnetic field. This is consistent with the notion that the orbital effects of a magnetic field are given by a Lorentz force, which is a vector product of the velocity with the field, thus leaving motion along the direction of the applied field unaffected. This agrees with the prediction of Eq.~\eqref{eq:dH}, giving that $\hat{x}_d \approx \hat{x}$, meaning that the dot is oriented along the device axes. The striking invariance of $E_y$ indicates that the corresponding orbital is rather well aligned with the magnetic field and therefore the $y$-axis of the device. In this case, the parallel axis drops out of the 3D description and the problem is reduced to a mixing of the $x$- and $z$-like orbital. Comparing the two cases in Fig.~\ref{fig:Fig2}, we emphasize that, going from panel (b) to (c), the quantum dot was in fact rotated by $90^\circ$, thus demonstrating a gate-induced quantum dot \textit{rotation}. Indeed, this is expected from the gate voltage dependence $V_{\rm shape}$, and is here validated in real space with the in-plane field spectroscopy.

By fitting the data to Eqs.~\eqref{eq:flux} and \eqref{eq:dE}, we can extract the effective width $\lambda_z$, and thus the size of the quantum dot along the growth direction. We can convert the latter, under a rather mild assumption that the heterostructure confinement is triangular, to the interface electric field $E_{\rm ext}$ and the subband energy splitting $E_z$. This in turn allows for the evaluation of the spin-orbit fields. Namely, from $\lambda_z=6.3\pm0.3$ nm we get the spin-orbit lengths $l_{\rm r}=2.1\pm0.3$ $\mathrm{\mu}$m and $l_{\rm d}=3.2\pm0.3$ $\mathrm{\mu}$m for Rashba and Dresselhaus interaction, respectively \cite{ErrorBars}. Using an independent fit from the directional variation of the spin-relaxation time, Ref.~\cite{Camenzind2017}, gave $l_{\rm r}=2.5\pm0.2$ $\mathrm{\mu}$m, and $l_{\rm d}=4.1\pm{0.4}$ $\mathrm{\mu}$m illustrating the agreement. We point out that, apart from determining the spin-orbit interactions strengths, the width of the 2DEG determines also the strength the electron Fermi-contact interaction with nuclear spins. Thus, knowledge on the quantum-dot size along the growth direction is essential for quantitative analysis of spin properties, such as relaxation \cite{Camenzind2017}.

\begin{figure}\vspace{-3mm}
\includegraphics[width=\columnwidth]{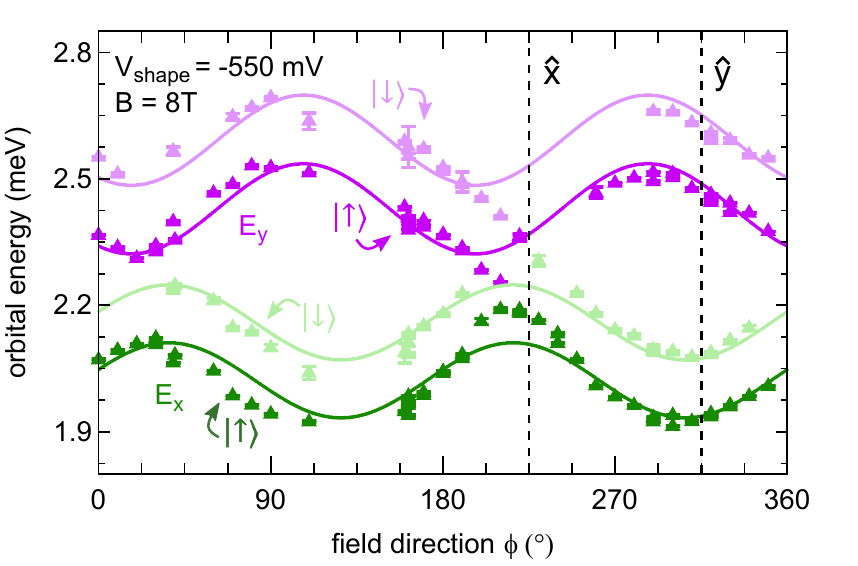}\vspace{-2mm}
\caption{Spin-resolved excitation-energies measured at magnetic field of a fixed magnitude $8$\,T and varying direction in an almost circular quantum dot ($V_{\rm shape}\sim-550$\,mV in Fig. \ref{fig:Fig2}). The solid lines show a fit according to Eq.~\eqref{eq:dE} for each orbital state (separately for the green and purple data) assuming a direction independent Zeeman energy. The fit gives 
$\tilde{\delta}_x=-8^\circ\pm4^\circ$, 
$\tilde{\delta}_y=62^\circ\pm4^\circ$ and $\lambda_z = 6.1\pm0.3$\,nm ($E_z=30.7\pm3$\,meV)\cite{GeneralizedFit}.}
\vspace{-5mm}
\label{fig:Fig4}
\end{figure}

We now turn to the quantification of the dot orientation. It can be done by measuring the excitation energies at a magnetic field with fixed magnitude and varied orientation. Figure \ref{fig:Fig4} presents such data for $B=8$\,T and a more symmetric dot. The energies show a sinusoidal behavior as expected from Eq.~\eqref{eq:dE}. The two states oscillate out of phase, proving that they represent orbitals oriented perpendicular to each other, see also \cite{Supp}S3. For an elongated (quasi-1D) dot, the states would oscillate in phase \cite{Stano2017}. Beyond confirming that our dot is indeed close to a symmetric one, we can specify its orientation in detail. By fitting the data of Fig.~\ref{fig:Fig4} to Eqs.~\eqref{eq:H} and \eqref{eq:dH}, we obtain $\tilde{\delta}=-8^\circ\pm4^\circ$, indicating the dot is only slightly tilted away from the device coordinate system. We note that even such modest misalignment can have large impact on the qubit quality \cite{Malkoc2016}, and on characterization of the spin-orbit fields \cite{Scarlino2014}.

Before concluding, we look at the assumption that the in-plane confinement is bi-quadratic, adopted in Eq.~\eqref{eq:H}. It has been used from the onset of quantum-dot investigations \cite{Reimann2002}, as a practical choice for which analytical solutions are known \cite{Schuh1985,Davies1985,Rebane1972}. Compared to its prevalent use, the evidence on such confinement shape is less abundant, and has been up to now restricted to checking the equidistant energy spacing of excited states of a harmonic oscillator. The data in Fig.~\ref{fig:Fig4} can provide additional information. Namely, fitting each of the accessible orbitals to Eq.~\eqref{eq:dE} {\it individually}, we can extract the x,y orbital-specific angle $\delta_{x,y}$. In principle, one can map-out the dependence of $\delta$ on the single-particle state energy, if more excited states are accessible. Here, we find $\delta_y-\delta_x \approx 70\pm8^\circ$, slightly different from $90^\circ$, thus deviating slightly from an ideal bi-harmonic potential, and here we have quantified by how much.

In summary, we measure excitation energies in a single-electron lateral quantum dot with in-plane magnetic fields of varying orientation. We show that such measurement can determine the orientation of the dot, and reconstruct its single-particle quantum-mechanical orbitals. In particular, this means that for a given orbital, one can assign a size and orientation within the 2DEG plane, as well as its extension along the growth direction with sub-nm resolution. The information on the quantum dot shape has an immediate use in correct quantification of the spin-orbit fields as well as the strength of the electron-nuclear Fermi contact hyperfine interaction. We note that the method is directly applicable to any quasi-2D dot, also in other materials, and more sophisticated structures, for example, triple-quantum-dot devices with non-collinear arrangement, as well as dots with higher electron occupations, where Hartree-Fock orbitals could be accessed in the same way.

We thank V. Golovach for valuable inputs and stimulating discussions and M. Steinacher and S. Martin for technical support. This work was supported by the Swiss Nanoscience Institute (SNI), NCCR QSIT, Swiss NSF, ERC starting grant (DMZ), and the European Microkelvin Platform (EMP). PS acknowledges support from CREST JST (JPMJCR1675), and JSPS Kakenhi Grant No.~16K05411.

\section*{References}

\begin{thebibliography}{36}
\bibitem [{Err()}]{ErrorBars}%
  \BibitemOpen
  \href@noop {} {}\bibinfo {note} {The error intervals on the spin-orbit
  lengths given here are due to the uncertainty in $\lambda_z$. We do not
  reflect additional uncertainty, stemming from the conversion factors
  (basically, uncertainty in parameters of the k.p theory).}\BibitemShut
  {Stop}%
\end{thebibliography}

%


\end{document}